\documentclass[aps,prl,twocolumn,reprint,amsmath,amssymb,superscriptaddress,floatfix]{revtex4-1}

\usepackage{helvet}
\usepackage[pdftex]{hyperref}
\hypersetup{pdftitle={Al6Re},pdfauthor={Erjian Cheng},pdfsubject={Al6Re},pdfdisplaydoctitle}
\usepackage{graphicx}
\usepackage{bm}

\def\Tc{\ensuremath{T_\text{c}}}
\def\Hc{\ensuremath{H_\text{c}}}

\begin{document}

\title{Magnetotransport in Al$_6$Re}

\author{Erjian Cheng}
\affiliation{State Key Laboratory of Surface Physics, Department of Physics, and Laboratory of Advanced Materials, Fudan University, Shanghai 200438, China}

\author{Darren C. Peets}
\email{dpeets@nimte.ac.cn}
\affiliation{State Key Laboratory of Surface Physics, Department of Physics, and Laboratory of Advanced Materials, Fudan University, Shanghai 200438, China}
\affiliation{Ningbo Institute of Materials Technology and Engineering, Chinese Academy of Sciences, Ningbo, Zhejiang 315201, China}

\author{Chuanying Xi}
\affiliation{Anhui Province Key Laboratory of Condensed Matter Physics at Extreme Conditions, High Magnetic Field Laboratory of the Chinese Academy of Sciences, Hefei, Anhui 230031, China}

\author{Yeyu Huang}
\affiliation{State Key Laboratory of Surface Physics, Department of Physics, and Laboratory of Advanced Materials, Fudan University, Shanghai 200438, China}

\author{Li Pi}
\affiliation{Anhui Province Key Laboratory of Condensed Matter Physics at Extreme Conditions, High Magnetic Field Laboratory of the Chinese Academy of Sciences, Hefei, Anhui 230031, China}

\author{Shiyan Li}
\email{shiyan\_li@fudan.edu.cn}
\affiliation{State Key Laboratory of Surface Physics, Department of Physics, and Laboratory of Advanced Materials, Fudan University, Shanghai 200438, China}
\affiliation{Collaborative Innovation Center of Advanced Microstructures, Nanjing 210093, China}

\begin{abstract}
  Since very few Type-I superconductors are known and most are elemental superconductors, there are very few experimental platforms where the interaction between Type-I superconductivity and topologically nontrivial band structure can be probed.  The rhenium aluminide Al$_6$Re has recently been identified as a Type-I superconductor with a transition of 0.74\,K and a critical field of $\sim$50\,Oe. Here, we report its magnetotransport behavior including de Haas-van Alphen (dHvA) and Shubnikov-de Haas (SdH) oscillations. Angular dependence of the magnetoresistance reveals a highly anisotropic Fermi surface with dominant hole character. From the strong oscillatory component $\Delta R_{xx}$ in high magnetic fields up to 33\,T, the Landau index infinite-field intercept in the case of a single oscillation frequency, and the phase factor $\varphi$ where multiple frequencies coexist, are both $\sim$1/4.  This intermediate value is suggestive of possible nontrivial band topology but does not allow strong conclusions.


\end{abstract}

\maketitle

\section{Introduction}

Topological materials have attracted tremendous interest in
condensed matter physics, since their low energy excitations are massless Dirac or Weyl fermions originally
introduced in high energy physics \cite{Reviews1,Reviews2}, and because their spin-split band dispersion offers high-fidelity spin transport with minimal scattering, which is expected to be indispensible for spintronics.
Such materials are characterized by the number
and the type of all of their band crossings \cite{Reviews1,Reviews2,TNL-2}.
One particularly interesting class of topological materials is topological semimetals, which can be further
classified into nodal-point and nodal-line semimetals
based on the degeneracy and dimensionality of their nodes, with effective Hamiltonians for Weyl and Dirac
fermions given by the Weyl and gapless-Dirac equations, respectively \cite{Weyl,Dirac}.
In Dirac semimetals, the conduction and valence bands cross
at unique points in the Brillouin zone (Dirac points) with linear energy dispersion. The Dirac node is fourfold degenerate
and can split into two doubly-degenerate Weyl nodes with opposite
chirality through symmetry breaking in momentum space \cite{Reviews1,Reviews2}.
Unlike the discrete points in Dirac or Weyl semimetals, the band crossings
in nodal-line semimetals form closed loops \cite{TNL-2}.
The topological semimetal family exhibits a wide range of exotic and potentially useful properties,
such as a high charge carrier mobility, large magnetoresistance (MR), and the chiral anomaly \cite{Reviews1,Reviews2,Dirac-2}.

\begin{figure}[htb]
\includegraphics[width=\columnwidth]{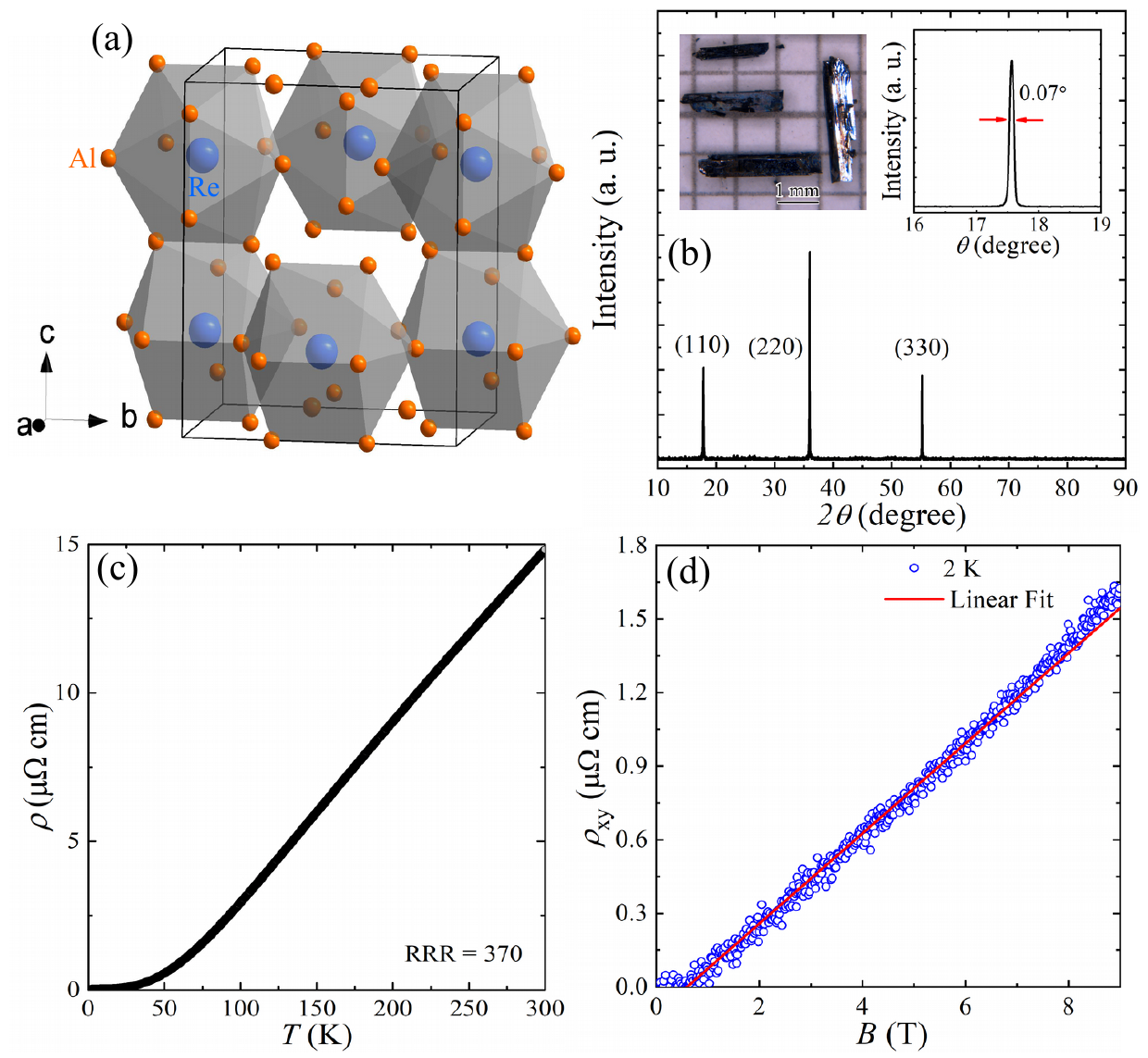}
\caption{\label{fig1}
(a) Crystal structure of Al$_6$Re. Blue and orange spheres represent Re and Al atoms, respectively. The Re atoms are located at the centres of Al cages.
(b) X-ray diffraction pattern from a large flat surface of the Al$_6$Re single crystal, identifying it as (110).
Left inset: optical micrograph of Al$_6$Re single crystals placed on a $1\times 1$\,mm$^2$ grid.
Right inset: X-ray rocking curve of the (220) Bragg peak with a full width at half maximum of 0.07$^\circ$.
(c) Longitudinal resistivity of Al$_6$Re in zero magnetic field.
(d) Transverse Hall resistivity of Al$_6$Re; the red line represents a linear fit.
}
\end{figure}

\begin{figure*}
\includegraphics[width=\textwidth]{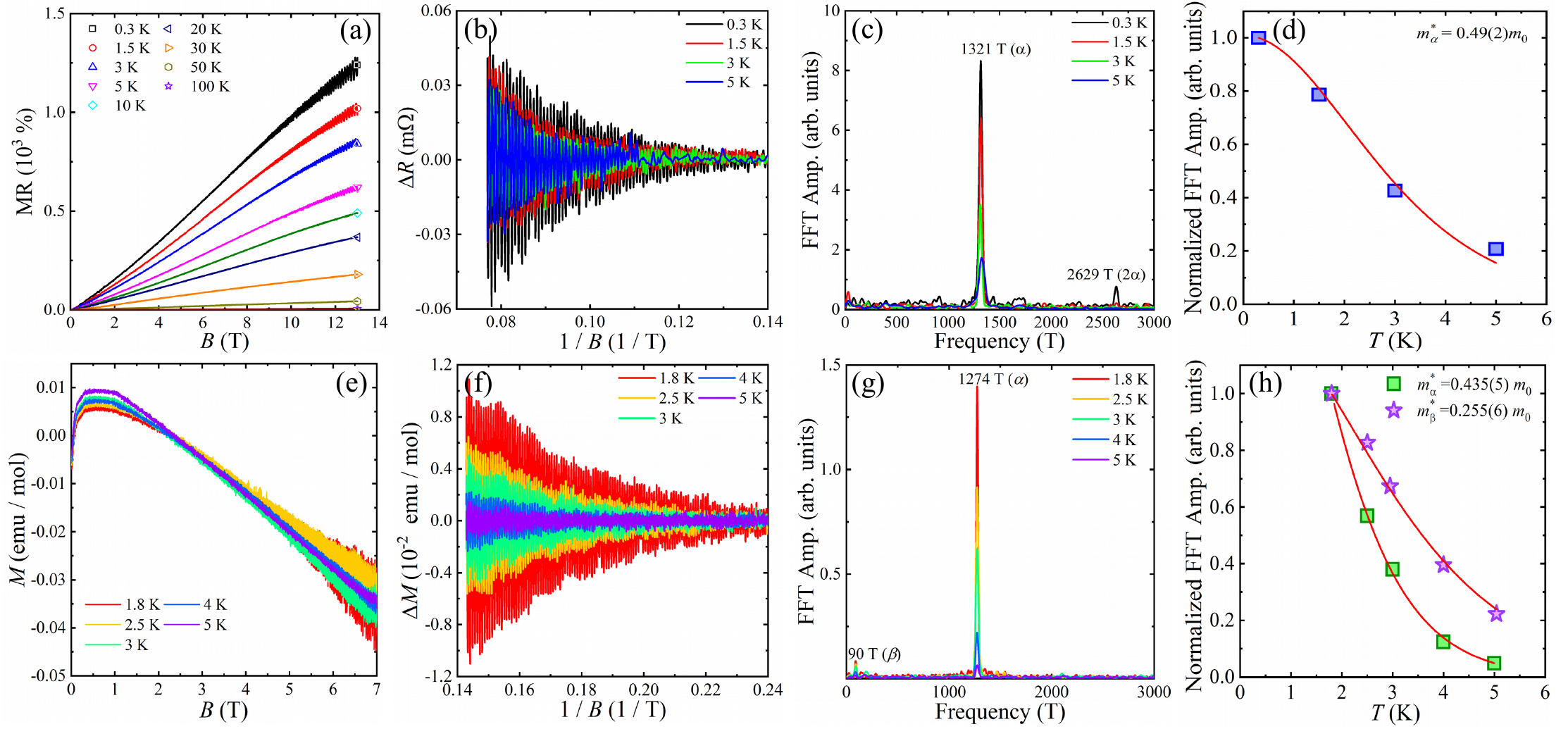}
\caption{\label{fig2}
Quantum oscillations in Al$_6$Re.
(a) Magnetoresistance at various temperatures, showing SdH oscillations.
(b) The oscillatory component $\Delta R_{xx}$, extracted from $R_{xx}$ by subtracting
a smooth background, as a function of 1/$B$ at various temperatures.
(c) FFT results for SdH oscillations at various temperatures.
(d) FFT amplitude {\itshape vs}.\ temperature for SdH.
The solid line represents the Lifshitz-Kosevich formula fit.
(e) dHvA oscillations of the magnetization at various temperatures.
(f) The oscillatory component $\Delta M$, extracted from $M$ by subtracting a
smooth background, as a function of 1/$B$ at various temperatures.
(g) FFT results for dHvA oscillations at various temperatures.
(h) FFT amplitude {\itshape vs}.\ temperature for dHvA.
The solid lines represent the Lifshitz-Kosevich formula fits.
}
\end{figure*}

The interactions of topologically nontrivial band structure with superconductivity, in topological superconductors (TSCs), is of
particular interest.  TSCs have a full gap in the bulk and Majorana
fermion states on the surface which can be used for high-performance
electronics and quantum computing \cite{TSC1,TSC2}.
Despite intense effort both in attempting to make topological materials superconduct and in identifying known superconductors that may have nontrivial band topology,
few topological superconductor candidates have been identified, the most notable ones being FeTe$_{0.55}$Se$_{0.45}$ \cite{FeTeSe-HDing}, Cu$_x$Bi$_2$Se$_3$ \cite{CuBiSe-DLFeng}, and (Li$_{0.84}$Fe$_{0.16}$)OHFeSe \cite {liu2018robust}.
These are typically Type-II superconductors, which can host additional Majorana modes inside vortex cores.
Candidate Type-I TSCs are exceedingly scarce.

Recently, we reported Type-I superconductivity in Al$_6$Re with a superconducting transition temperature \Tc\ of 0.74\,K
and a critical field \Hc\ of $\sim$50\,Oe \cite{Al6Re-SC}. Magnetization,
ac susceptibility, and specific heat data identified Al$_6$Re as a
Type-I superconductor.
Most of the pure elemental superconductors are Type-I, but only around a dozen
superconducting compounds are known to be Type-I.  Having identified Al$_6$Re as
a rare Type-I superconductor, we turn to the question of whether any topologically-nontrivial band features could make it an even rarer example of a Type-I topological superconductor. Recently, Zhang {\itshape et al}. reported an
effective and fully automated algorithm to sweep through the published crystal structures of all non-magnetic
materials, obtaining their topological invariants, and 8056 materials were suggested to
possess nontrivial band topology \cite{C.Fang2019}. The resulting online band structure calculation database posits that
Al$_6$Re should be a high-symmetry-point semimetal (HSPSM) with relevant band crossings at the $R$,
$T$ and $\Gamma$ points \cite{Al6Re-band}. The addition of spin-orbit coupling changes the topological invariants, but not the classification of Al$_6$Re as a HSPSM, nor the high-symmetry points at which the band crossings occur.
To check whether Al$_6$Re is topologically nontrivial, we performed magnetotransport measurements
on Al$_6$Re single crystals.

In this paper, we report de Haas-van Alphen (dHvA) and Shubnikov-de Haas (SdH) quantum oscillations on Al$_6$Re. A magnetoresistance of 3000\,\%\ at 33\,T has been observed, which is larger
than that for conventional metals \cite{Pippard1989}, while its angular dependence
reveals a large and extremely anisotropic Fermi surface (FS).
From the strong oscillatory component $\Delta R_{xx}$,
we obtained the linear dependence of the Landau index $n$ on 1/$B$,
and the Berry phase is further analyzed.

\section{Experimental}

Single crystals of Al$_6$Re were grown from Al flux as described in
detail elsewhere \cite{Al6Re-SC}. The as-grown single crystals are
hollow square tubes [see the left inset in Fig.~\ref{fig1}(b)]. For electrical transport
and magnetic susceptibility measurements, a hollow single crystal was
split to obtain a single isolated side, and then
the sample was cut into a bar shape $1.0\times 0.6\times 0.05$\,mm$^3$ in size for electric transport.
A standard four-probe method was used for the longitudinal
resistivity measurement, with the current applied in the (110) plane.
Data at 0.3\,K and above 1.5\,K were collected in a $^3$He and a $^4$He cryostat, respectively.
Magnetic susceptibility measurements were performed in a Quantum Design magnetic property
measurement system (MPMS, Quantum Design). High-field measurements were performed at the Steady
High Magnetic Field Facilities, High Magnetic Field Laboratory, Chinese Academy of Sciences, in Hefei.

\begin{figure*}
\includegraphics[width=\textwidth]{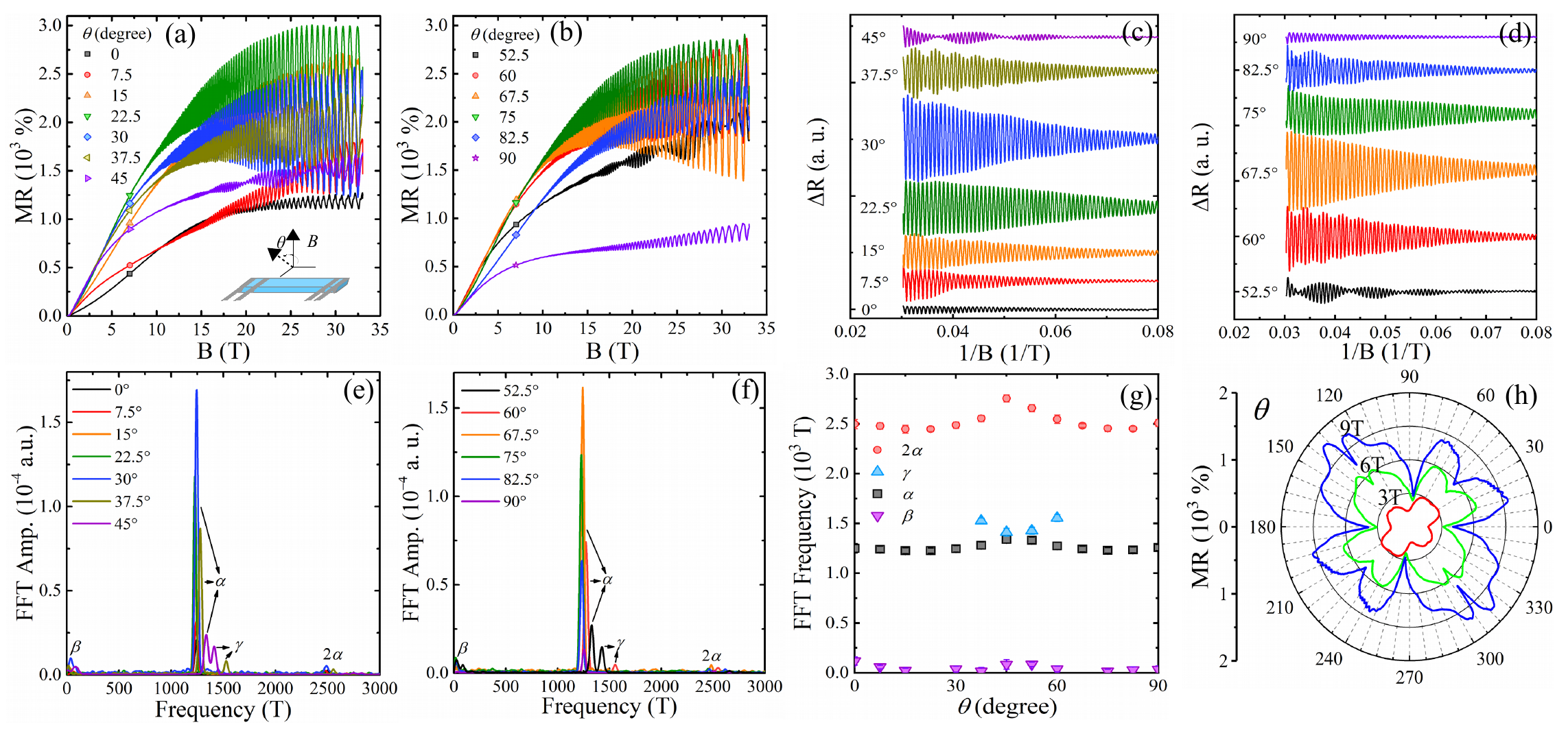}
\caption{\label{fig3}
Angular dependence of quantum oscillations.
(a,b) SdH oscillation of magnetoresistance;
the inset is a schematic illustration of the experimental geometry and the angle $\theta$.
(c,d) The oscillatory components $\Delta R_{xx}$ as a function of 1/$B$.
(e,f) FFT results.
The angular dependence of these frequencies is plotted in (g), where error bars represent the full widths at half maximum of the FFT peaks.
(h) Polar plot of the magnetoresistance in several magnetic fields.
}
\end{figure*}

\section{Results and Discussion}

Figure \ref{fig1}(a) shows the crystal structure of Al$_6$Re. The material has a MnAl$_6$-type structure with
space group ${Cmcm}$ (\#~63), and each rhenium atom has ten aluminum neighbors \cite{ReAl6-1}.
Figure \ref{fig1}(b) shows the X-ray powder diffraction pattern from a large flat side of the Al$_6$Re single crystal --- only ($hh0$) Bragg peaks appear. The full width at half maximum (FWHM) of the (220) peak
is 0.07$^\circ$ (right inset), indicative of the high quality of our Al$_6$Re single crystal.
Figure \ref{fig1}(c) presents the longitudinal resistivity of a Al$_6$Re  single crystal in zero
magnetic field. The residual resistivity ratio (RRR) $\rho_{300\,\text{K}}/\rho_0$ = 370 with
$\rho_0$ = 0.04 $\mu\Omega$cm, where the low residual contribution and large RRR further demonstrate the high quality of our
single crystal.

Figure \ref{fig1}(d) plots the transverse Hall resistivity of Al$_6$Re at 2\,K.
These data are more noisy due to the small absolute value of the Hall resistivity and the instrumental resolution, but it can be seen to exhibit a roughly linear dependence.
The slope of the Hall resistivity is positive, indicating that hole carriers dominate the transport.
An approximate linear fit to the Hall resistivity, respresented by the red line in Fig.~\ref{fig1}(d), yields an estimated carrier mobility of
$2.0\times 10^3$\,cm$^2$V$^{-1}$s$^{-1}$ and a carrier concentration of
$3.4\times 10^{21}$\,cm$^{-3}$. For the topological Dirac semimetal Cd$_3$As$_2$,
the dominant charge carriers are electronlike, and the carrier concentration is
$5.3\times 10^{18}$\,cm$^{-3}$ while the mobility is $4.1\times 10^4$\,cm$^2$V$^{-1}$s$^{-1}$ \cite{Cd3As2-LPHe}.
The much higher carrier concentration in Al$_6$Re indicates that Al$_6$Re is a good metal with substantial electron density at the Fermi level, independent of its topological properties.

In topological semimetals, a large MR and $\pi$ Berry phase
are generally observed \cite{Reviews1}, and these are often taken as strong evidence suggesting topologically nontrivial band structure. As shown in Fig.~\ref{fig2}(a), Al$_6$Re has an unsaturated MR of $\sim$1250\,\%\
at 13\,T and 0.3\,K with an evident SdH oscillation. On warming, the MR is gradually suppressed.
In our previous report, we found evidence that the superconducting critical current $J_c$ of Al$_6$Re
is extremely low \cite{Al6Re-SC}, and the excitation of 1\,mA used here is sufficient to completely squelch the superconductivity,
{\itshape i.e}.\ the MR measurements were conducted exclusively in the normal state.
We extracted
the oscillatory component $\Delta R_{xx}$ [Fig.~\ref{fig2}(b)] from $R_{xx}$ by subtracting a smooth
background. By analyzing the oscillatory component via fast Fourier
transform (FFT), frequency components at $F$ = 1321 and 2629\,T
were identified, the latter being the second harmonic of the former.
The band having a frequency of 1321\,T is referred to here as the $\alpha$ band.
Employing the Onsager relation $F$ = ($\hbar/2\pi e)A_F$,
the extremal cross-sectional area $A_F$ of the $\alpha$ band's Fermi surface is determined to be a relatively large 0.127\,\AA$^{-2}$.

The SdH oscillation amplitude can be described by the Lifshitz-Kosevich (LK) formula
\cite{Cd3As2-LPHe,LK_formula-1,LK_formula-2}:
\begin{displaymath}
\Delta\rho_{xx}\propto\frac{2\pi^2k_BT/\hbar\omega_c}{\sinh(2\pi^2k_BT/\hbar\omega_c)}e^{-\frac{2\pi^2k_BT_D}{\hbar\omega_c}}\cos2\pi\left(\frac{F}{B}+\frac{1}{2}+\beta\right)\text{,}
\end{displaymath}
where $\omega_c$ = $eB/{m^*}$ is the cyclotron frequency and $T_D$ is the Dingle temperature.
The Berry phase 2$\pi\beta$ will be discussed later.
The energy gap $\hbar\omega_c$ can be obtained from the
thermal damping factor $R_T$ = $2\pi^2k_BT/\hbar\omega_c/sinh(2\pi^2k_BT/\hbar\omega_c)$.
The best fit to the data, shown in Fig.~\ref{fig2}(d), yields $\hbar\omega_c$ $\approx$ 2.18\,meV.
The cyclotron effective mass $m^*$ of Al$_6$Re can be obtained through this fit, resulting in $m^*$ = 0.49(2)\,$m_0$.

We also checked for quantum oscillations in the magnetic susceptibility of our Al$_6$Re single crystal ---
Fig.~\ref{fig2}(e) plots the dHvA oscillations of Al$_6$Re at various temperatures.
Figure~\ref{fig2}(f) shows the oscillatory component of the magnetization, extracted by subtracting a smooth background as was done for the SdH oscillations.
A fast Fourier transform revealed
two oscillation frequencies --- 90 and 1274\,T.  The observation of an additional frequency in dHvA, which we refer to as the $\beta$ band, may be due to the anisotropic
Fermi surface of Al$_6$Re.
The magnetization oscillations in a Dirac system can also be described by
the LK formula \cite{ZrSiSPRL}, with the effective mass $m^*$ again obtainable through a fit
of the temperature dependence of the oscillation amplitude to the thermal damping factor $R_T$,
as shown in Fig.~\ref{fig2}(h). The obtained
effective masses are 0.255(6) and 0.435(5)\,$m_0$ for 90 and 1274\,T, respectively, and we note that the values from SdH and dHvA for the $\alpha$ band are in reasonable agreement.
The cross-sectional areas of the corresponding Fermi surfaces are determined to be
0.0086 and 0.122\,\AA$^{-2}$ for the 90 and 1274\,T oscillations,
respectively, indicating a complicated Fermi surface. Note that the obtained effective masses are heavier than some topological semimetals such as Cd$_3$As$_2$ ($\sim$ 0.044 $m_0$) \cite{Cd3As2-LPHe}, but lighter than some others such as PtBi$_2$ ($\sim$ 0.68 $m_0$) \cite{PtBi2theory,PtBi2experiment}.

To better reveal the Fermi surface anisotropy of Al$_6$Re, we performed angle-dependent
MR measurements at 1.5 K. Here, $\theta$ represents the angle between magnetic field $B$ and the [110] axis, as drawn schematically in the inset of Fig.~\ref{fig3}(a).
The MR is clearly anisotropic, but is saturated under higher magnetic field.
Around 45$^\circ$, an obvious additional oscillation frequency emerges, which we refer to as $\gamma$.
Figures~\ref{fig3}(c) and (d) show the SdH oscillations at different angles
after subtracting a smooth background.  SdH oscillations can be
observed at all angles but with varying frequency and intensity, indicating an anisotropic three-dimensional (3D) Fermi surface. We extracted the FFT
frequencies at all angles [Figs.~\ref{fig3}(e) and (f)], and the angular dependence of all frequencies
is plotted in Fig.~\ref{fig3}(g). We find that the new oscillation
frequency $\gamma$ branches off a cusp in $\alpha$. The frequencies have a
nonmonotonic angular dependence, indicating the presence of complex Fermi surface contours.
We cannot fit the angular dependence of the frequencies using a standard 2D
or 3D ellipsoidal Fermi surface model, further indicating that the Fermi surface of Al$_6$Re
is quite complex. Figure~\ref{fig3}(h) presents the angular dependence of the isothermal MR,
which exhibits a ``butterfly'' shape. This pattern exhibits twofold rotational symmetry as expected for an orthorhombic lattice, but the $Cmcm$ space group also has mirror planes which are not reflected here.

\begin{figure}[hbt]
\includegraphics[width=\columnwidth]{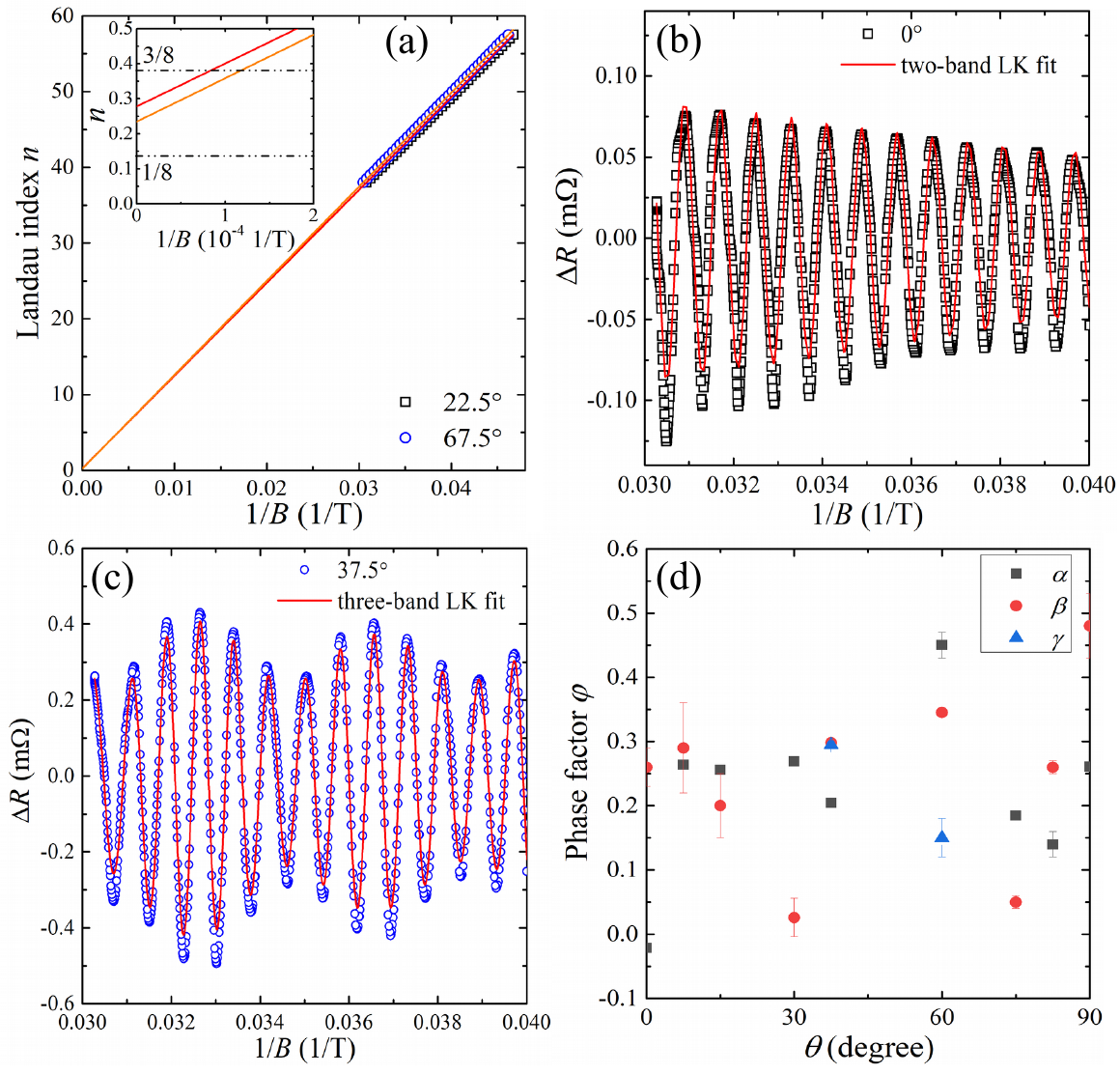}
\caption{\label{fig4}
Berry phase in Al$_6$Re. (a) Landau index $n$ plotted against 1/$B$ for the SdH oscillations at 22.5$^\circ$ and 67.5$^\circ$. Lines represent linear fits. Inset shows the extrapolation of 1/$B$ to zero --- the intercepts lie around 1/4. (b) and (c) show SdH oscillations at 0$^\circ$ and 37.5$^\circ$, respectively. The solid red lines represent fits to multiple Lifshitz-Kosevich functions \cite{ZrSiSPRB,ZrSiSPRL,pavlosiuk2016giant}. (d) Angle dependence of the phase factor $\varphi$ as discussed in Ref.\cite{pavlosiuk2016giant}.
}
\end{figure}

We now return to the Berry phase.
A nontrivial Berry phase is generally considered to be key evidence for Dirac fermions and has been observed in other Dirac materials.
For those topological semimetals with a single oscillation frequency, for example Cd$_3$As$_2$ \cite{Cd3As2-LPHe}, the Landau index can be directly extracted by linearly fitting the peaks and valleys in $\Delta R_{xx}$ as a function of 1/$B$ in the Landau-level fan diagram. However, for multi-frequency topological semimetals, for example ZrSi$X$ ($X$ = S, Se, Te) \cite {ZrSiSPRB,ZrSiSPRL}, the maxima of $\Delta R_{xx}$ do not correspond to the maxima of particular components with different frequencies, so it is inappropriate to determine the Berry phase in a multi-component oscillation through the Landau-level fan diagram.
According to the angle-dependent MR measurements shown in Fig.~\ref{fig3}, there are only a single frequency ($\alpha$) and its harmonic (2$\alpha$) at 22.5$^\circ$ and 67.5$^\circ$.
Figure~\ref{fig4}(a) displays Landau index plots for the MR at 22.5$^\circ$ and 67.5$^\circ$.
At this angle, we can assign integer indices to the $\Delta R_{xx}$ peak positions
in 1/$B$ and half-integer indices to the $\Delta R_{xx}$ valleys.
According to the Lifshitz-Onsager quantization rule $A_F(\hbar/eB)$ =
$2\pi(n+1/2+\beta+\delta)$, the Landau index $n$ is linearly dependent
on 1/$B$. $2\pi\beta$ is the Berry phase, and $2\pi\delta$ is an additional
phase shift resulting from the curvature of the Fermi surface in the third dimension.
$\delta$ varies from 0 for a quasi-2D cylindrical Fermi surface to $\pm$ 1/8 for a corrugated 3D Fermi surface, thus values around $-$1/8 to 1/8 are generally associated with nontrivial band topology in 3D, while intercepts around 3/8 to 5/8 indicate trivial band topology.
Our data points in Fig.~\ref{fig4}(a) fall into very straight lines allowing reliable extrapolations to infinite field, and the intercepts of $0.27\pm0.02$ for 22.5$^\circ$ and $0.23\pm0.02$ for 67.5$^\circ$ fall in the range between 1/8 and 3/8, as shown in the inset of Fig.~\ref{fig4}(a).
These intermediate values are suggestive of a nontrivial Berry phase, but do not allow a strong conclusion.
Since the Fermi surface of Al$_6$Re is quite large, the lowest Landau level accessible in our
experiments is just under 40, so extension of this study to higher magnetic field would lead to a more accurate value for the intercept.

Apart from the Landau index fan diagram, the Berry phase can also be determined directly from a fit to the LK formula. A multiple-LK function can be used to fit the multi-component oscillation \cite{ZrSiSPRB,ZrSiSPRL,pavlosiuk2016giant}.
Figures~\ref{fig4}(b) and (c) show the SdH oscillations at 0$^\circ$ and 37.5$^\circ$ fit using a two-band and a three-band LK formula, respectively. As one can see, the data are very well fit. The fit at 0$^\circ$ yields a phase factor $\varphi$ \cite{pavlosiuk2016giant} of $-$0.021$\pm$0.001 for the $\alpha$ band and 0.26$\pm$0.03 for the $\beta$ band. At 37.5$^\circ$, we obtain phase factors of 0.205$\pm$0.001, 0.298$\pm$0.005, and 0.295$\pm$0.002 for the $\alpha$, $\beta$, and $\gamma$ bands, respectively. For free electrons, a phase factor close to the Onsager expectation of 1/2 suggests no Berry phase, while a phase factor close to zero is indicative of a nontrivial $\pi$ Berry phase \cite {pavlosiuk2016giant}. We checked the remaining angles using multiple LK formulas and excellent fits were obtained for all angles other than 45 and 52.5$^\circ$, where the near-degeneracy of the $\alpha$ and $\gamma$ freqencies proved problematic. Figure~\ref{fig4}(d) plots the resulting angle dependence of the phase factor $\varphi$ for all three frequencies. The values are dispersed over a wide range around 0.2. The average values and their standard deviations are 0.22$\pm$0.13, 0.25$\pm$0.14, and 0.22$\pm$0.10 for the $\alpha$, $\beta$ and $\gamma$ bands, respectively. While values that differ significantly from 1/2 are suggestive of a deviation from topologically trivial expectations, and our values are marginally closer to 0 than to 1/2, these intermediate values and their scatter do not allow a strong conclusion as to whether Al$_6$Re possesses a non-trivial Berry phase. Finally, we note that LK fits are often able to provide other parameters, such as the Dingle temperature, but in this case the uncertainties are even larger on these other quantities. As for the angles with only one frequency, measuring to higher fields would help. Additional experimental evidence, such as from angular-resolved photoemission or scanning tunneling spectroscopy, would also be very useful in resolving the behavior of Al$_6$Re.

As Zhang {\itshape et al}.\ pointed out \cite{C.Fang2019}, conventional metals can also be HSPSMs, but these usually have topologically trivial band structures. However, conventional metals seldom exhibit MR as large as that in Al$_6$Re. A family of transition-metal icosagenides with a body-centered tetragonal lattice, including the Al-based alloy VAl$_3$, was proposed to be Lorentz-violating Type-II Dirac semimetals \cite{Tay-Rong_Chang} with a unique Landau level spectrum. Although this proposal remains controversial \cite{VAl3-1,VAl3-2}, it still provides a useful point of reference for
topology in Al-based alloys, including Al$_6$Re. Landau index intercepts of 1/8 and 3/8 were observed for different bands in VAl$_3$ \cite{VAl3-1}, but a topologically trivial explanation was possible for both values. In the case of Al$_6$Re, we cannot exclude the possibility that Berry phase values around 0.2 can also be explained in terms of topologically trivial band structure.

\section{Conclusion}

In summary, we report unusual magnetotransport behavior in Al$_6$Re by means of de Haas-van Alphen (dHvA) and Shubnikov-de Hass (SdH) oscillations and angular-dependent magnetoresistance, revealing signs of a large, complex, and anisotropic Fermi surface.
The Landau index intercepts and phase factors fall in an intermediate range around 1/4, which is inconsistent with the expectations for a topologically trivial 3D Fermi surface, suggestive of nontrivial band topology. However, to conclusively determine whether Al$_6$Re is topologically non-trivial will require additional theoretical and experimental evidence.  Since Al$_6$Re may offer a unique platform for investigating the interaction of topology with Type-I superconductivity, in which Majorana fermions would be constrained to the surface, fully determining the topological properties of this material should be a particular priority.

\begin{acknowledgments}
This work is supported by the Ministry of Science and Technology of China (Grants No.\ 2015CB921401 and 2016YFA0300503), the National Natural Science Foundation of China (Grants No.\ 11674367 and 11650110428), the NSAF (Grant No.\ U1630248), and the Zhejiang Provincial Natural Science Foundation (Grant No.\ LZ18A040002). DCP is supported by the Chinese Academy of Sciences through 2018PM0036.
CYX is supported by the Users with Excellence Project of Hefei Science Center CAS through 2018HSC-UE015.
\end{acknowledgments}

\end{document}